\documentclass[iop]{emulateapj}
\usepackage{latexsym}
\usepackage{dcolumn}
\usepackage{bm}
\usepackage{amsmath}
\usepackage{natbib}
\input{epsf}
\newcommand{\be}{\begin{equation}}
\newcommand{\ee}{\end{equation}}
\newcommand{\ba}{\begin{eqnarray}}
\newcommand{\ea}{\end{eqnarray}}
\newcommand{\bi}{\begin{itemize}}
\newcommand{\ei}{\end{itemize}}
\newcommand{\bfi}{\begin{figure}[t]
\epsfxsize=9cm
\epsffile}
\newcommand{\bfig}{\begin{figure*}[t]
\epsfxsize=15cm
\epsffile}
\newcommand{\efi}{\end{figure}}
\newcommand{\efig}{\end{figure*}}
\newcommand{\no}{\nonumber}

\newcommand{\hmpc}{h/{\rm Mpc}}
\begin{document}
\title{A possible explanation of vanishing halo velocity bias}
\author{Pengjie
  Zhang$^{1,2,3,4}$\\
$^1$Department of Astronomy, School of Physics and Astronomy, Shanghai
Jiao Tong University\\ 
$^2$ IFSA Collaborative Innovation Center, Shanghai Jiao Tong
University, Shanghai 200240, China\\
$^3$ Tsung-Dao Lee Institute, Shanghai 200240, China\\
$^4$ Shanghai Key Laboratory for Particle Physics and Cosmology, Shanghai 200240, China
}
\email[Email me at: ]{zhangpj@sjtu.edu.cn}
\begin{abstract}
Recently Chen et al. (2018, ApJ, 861, 58) accurately determined the volume weighted halo velocity bias in
simulations, and found that the deviation of velocity bias from unity
is much weaker than the peak model prediction. Here we present a
possible explanation of this vanishing velocity bias. The starting
point is that, halos are peaks
in the low redshift {\it non-Gaussian} density field with smoothing scale
$R_{\Delta}$ (virial radius), instead of peaks in the high redshift
initial {\it Gaussian} density field with a factor of
$\mathcal{O}(\Delta^{1/3})$ larger smoothing scale. 
Based on the approximation that the density field can be Gaussianized by a local and monotonic
transformation, we extend the peak model to the non-Gaussian density field and derive the analytical expression
of velocity dispersion and velocity power spectrum of these
halos. The predicted deviation of velocity bias from unity is indeed
much weaker than the previous prediction, and the agreement with the
simulation results is significantly improved. 
\end{abstract}
\maketitle

\section{introduction}
The volume weighted halo/galaxy velocity bias at large scale ($\ga 10$ Mpc) is a long standing problem in
modern cosmology. It is not only of theoretical importance in
understanding the structure formation of the universe, but also of
practical importance in constraining dark energy with peculiar
velocity (e.g. redshift space distortion). Recently \citet{Chen18}
managed to measure the halo velocity bias in simulations, with
$0.1\%$-$1\%$ accuracy. This is achieved by a novel method, which
circumvents the sampling artifact problem
\citep{DTFE96,Bernardeau97,DTFE00}) prohibiting accurate 
velocity measurement \citep{Pueblas09,Zheng13,Zhang15a,Zheng15a}. A
major finding is that the deviation of velocity bias $b_v$ from unity
($|b_v-1|$) is very weak at $k\sim 0.1\hmpc$. For examples, all $z=0$ halos in the
mass rangle $5\times 10^{11}<M/(M_\odot/h)<10^{14}$ have  $|b_v-1|\la
0.1\%$ at $k\leq 0.1\hmpc$ (Table 2, \citet{Chen18}). This
finding validates the usual assumption of $b_v=1$ in data analysis
of peculiar velocity cosmology and eliminates a potential systematic
error associated with the velocity bias. 

However,  these numerical findings disagree with the
theoretical prediction of the peak model \citep{BBKS, Desjacques10}. The
peak model is based
on the correspondence of halos with peaks in 
the initial density field and the existence of correlation between
density gradient and velocity. Based on the two facts, the seminal
BBKS paper \citep{BBKS} predicted $\sigma_{v,{\rm 
    halo}}^2/\sigma_{v,{\rm matter}}^2<1$ and the deviation reaches $10\%$
for $10^{13}M_\odot$ halos. \citet{Desjacques10} (hereafter DS10)
extended BBKS to 2-point statistics, and derived an elegant
expression $b_v(k)=1-R_v^2(M)k^2$.  For $10^{13}M_\odot$ halos, the
deviation from unity is $\sim 5\%$ at $k=0.1\hmpc$ and larger at
smaller scales. The peak model predictions (BBKS and DS10 to leading
order) are theoretically exact, in the context of proto-halos defined in the linear and
Gaussian initial conditions. Furthermore, the DS10 prediction has been verified in
N-body simulations \citep{Elia12,
  2015PhRvD..92l3507B}, and supported by further theoretical discussions
\citep{2012PhRvD..85h3509C,2014PhRvD..90j3529B,2015PhRvD..92l3525C}. 

The discrepancies between the peak model prediction and the
\citet{Chen18} numerical finding then require explanations. (1) The
difference in the halo/proto-halo definitions is likely the dominant
cause.  \citet{Chen18} adopted the usual definition of halos, identified
by the Friends-of-Friends (FoF) algorithm with linking length $b=0.2$
at the investigated redshifts. These halos correspond to peaks  in the 
late epoch non-Gaussian density field, smoothed with scale of the halo virial
radius $R_\Delta$. Here $\Delta\sim \mathcal{O}(100)$ is the mean halo density within
radius $R_\Delta$, in term of the mean cosmological matter density. On the other hand,
theoretical works of  the peak model  focus on ``proto-halos'', which
are peaks in the initial Gaussian density field, smoothed with scale
$R_S=\Delta^{1/3}R_\Delta$. Numerical works of the peak model  \citep{Elia12,
  2015PhRvD..92l3507B}  adopt a different definition of
``proto-halos'', as groups of particles which are members of $z=0$ 
  halos. Halos defined in the first way are hosts of galaxies in
  astronomical surveys, and therefore are directly 
related to the interpretation of galaxy velocity measurement. (2) Another difference
is that \citet{Chen18} defined the velocity bias as the ratio of halo
velocity and matter velocity at the same epoch
(e.g. $z=0,1,2$). It is directly related to the galaxy velocity
measurement at these redshifts, and its applications in measuring  the
structure growth rate and  constraining gravity (e.g. \citet{2013MNRAS.428..743L}). On the other hand, \citet{Elia12} only measured the velocity bias at
the initial redshifts ($z=50,70$), and  \citet{2015PhRvD..92l3507B}
defined the velocity bias with respect to the linearly evolved matter
velocity. Furthermore, the velocith bias measured by \citet{Chen18} is
volume weighted, while that by \citet{2015PhRvD..92l3507B} is (halo
number density) weighted. 

Motivated by  these possibilities, we present a quantitative derivation of
the volume weighted halo velocity bias at late epoch.  What we need is the
non-Gaussian joint PDF of the density, density gradient and velocity fields with smoothing scale
$R_\Delta$. Due to complexities in the nonlinear evolution, no exact
analytical expression exists. However, due to the fact that the
non-Gaussian density field can be Gaussianized to a good approximation
\citep{1991MNRAS.248....1C,1994ApJ...420...44K,2000MNRAS.314...92T,2001ApJ...561...22K,2009ApJ...698L..90N,Copula10,2010PhRvL.105y1301S,Yu11,2011ApJ...731..116N,2011PhRvD..83b3501S},
we are able to write down the joint PDF analytically. This allows us
to capture major impact of the nonlinear 
evolution on the halo velocity bias, and provides a possible
explanation on the observed $b_v\simeq 1$. 

\section{The velocity bias  in Gaussianized  field}
The N-point joint PDF of density field is fully captured by the
one-point PDF and the N-point copula \citep{Copula10}. The Copula
function is invariant under local and monotonic transformation of
density field. \citet{Copula10} found in N-body simulations that,
despite significant non-Gaussianity in the one-point PDF, the
two-point Copula is nearly Gaussian.  This means that, once we
perform a local transformation to render the one-point PDF Gaussian, the
two-point PDF will be Gaussian as well. The well-known lognormal
transformation
\citep{1991MNRAS.248....1C,1994ApJ...420...44K,2000MNRAS.314...92T,2001ApJ...561...22K,2009ApJ...698L..90N}
is an approximation to the local Gaussianization transformation. 

We will work under this {\it Gaussian Copula hypothesis}. We denote
the Gaussianization transformation as  $G=G(\delta)$, where $\delta$ is
the matter overdensity and $G$ is the Gaussianized field. By the
construction, both the one-point PDF $P(G)$ and the two-point PDF
$P(G_1,G_2)$ are Gaussian. The velocity field may have non-negligible non-Gaussian
component at small scale (e.g. internal motions within halos). Fortunately, the halo velocity smoothes and
suppresses the non-Gaussian velocity
components below scale of the halo virial radius. Furthermore, we are interested in the large scale
velocity statistics. Therefore we neglect non-Gaussianities in the
velocity field.  We are then able to write down the non-Gaussian joint
PDF of ${\bf v}$, $\delta$, and the density gradient $\nabla
\delta$. It is essential to include the density gradient in the joint
PDF, since it is correlated with the velocity field. Halos only reside at regions of
zero density gradient (density peaks), resulting in the velocity
bias. 

Since the joint PDF of $G$, $\nabla G$ and ${\bf v}$ is Gaussian, the
expression of velocity bias is identical to that of BBKS and DS10, once we
replace the initial linear density $\delta_L$ by $G$, $\nabla \delta_L$ by $\nabla G$, and the
smoothing scale $R_S=(3M/(4\pi \bar{\rho}_m))^{1/3}$ by
$R_\Delta=(3M/(4\pi \Delta \bar{\rho}_m))^{1/3}$. Here $\bar{\rho}_m$
is the (comoving) mean cosmological matter density. At high redshift,
$\Delta \rightarrow 178$. At redshift zero, $\Delta \simeq
100\rho_c/\bar{\rho}_m\simeq 350$ due to
the non-zero cosmological constant \citep{Eke96}. Extending BBKS to
the Gaussianized $G$ field, we obtain
\ba
\label{eqn:r2}
\frac{\sigma_{v_h}^2}{\sigma_v^2}=1-r^2\
, \ \ r^2\equiv \frac{\langle {\bf v}\cdot \nabla
  G\rangle^2}{\sigma_v^2\sigma_{\nabla G}^2}\ .
\ea
Extending DS10 to the $G$ field, we obtain 
\ba
\label{eqn:bv}
b_v(k)\simeq 1-R_v^2k^2\ ,\ R^2_v\equiv \frac{\sigma_G^2}{\sigma_{\nabla G}^2}\ .
\ea
Here $\sigma^2_\alpha\equiv \langle \alpha^2\rangle$
($\alpha=\delta,\nabla \delta, G, \nabla G, {\bf v}, {\bf v}_h$). We
emphasize again that all properties are the smoothed properties with
smoothing scale $R_\Delta$. 

For heuristic purpose, we show the derivation of Eq. \ref{eqn:r2}
here. 
The full derivation should be done in 3D, which is lengthy. However, as found in
BBKS and DS10, the expression of velocity bias in 1D can be converted
into the realistic 3D case straightforwardly. Therefore we will
only briefly present the derivation of 1D case, where the 3D
gradient $\nabla$ is replaced by the  1D $^{'}\equiv d/dx$. The joint
one-point PDF is 
\ba
P(v,\delta,\delta^{'})=P(v,G,G^{'})\left(\frac{dG}{d\delta}\right)^2\ .
\ea
Since $\langle G G^{'}\rangle=0$ and $\langle G
v\rangle=0$, the Gaussian PDF $P(v,G,G^{'})$ is separable
($P(v,G,G^{'})=P(G)P(v,G^{'})$). The relevant PDF for the velocity bias is 
\ba
P(v,G^{'})&=&\frac{1}{2\pi
  \sqrt{|{\bf C}|}}
\exp\left[-\frac{1}{2}\left(v^2(C^{-1})_{11}+\right. \right.\no \\
&&\left. \left. (G^{'})^2(C^{-1})_{22}+2vG^{'}(C^{-1})_{12}\right)
\right]\ .
\ea
The covariance matrix between $v$ and $G^{'}$ is 
\begin{displaymath}
\mathbf{C} =
\left( \begin{array}{cc}
C_{11} & C_{12} \\
C_{21} & C_{22} 
\end{array} \right)=
\left( \begin{array}{cc}
\sigma_v^2 & \langle vG^{'}\rangle \\
\langle vG^{'}\rangle& \sigma_{G^{'}}^2 
\end{array} \right)
\end{displaymath}
Halos satisfy $\delta=\Delta$ and $\delta^{'}=0$, and therefore
$G=G(\Delta)$ and $G^{'}=0$. The halo velocity
dispersion is 
\ba
\sigma_{v_h}^2&=&\frac{\int v^2
  P(v,\delta=\Delta,\delta^{'}=0)dv}{\int
  P(v,\delta=\Delta,\delta^{'}=0)dv}\no \\
&=&\frac{\int v^2
  P(v,G^{'}=0)dv}{\int P(v,G^{'}=0)dv}=\sigma_v^2\left(1-\frac{\langle
  vG^{'}\rangle^2}{\sigma_v^2\sigma_{G^{'}}}\right)\ .
\ea
Replacing the 1D gradient $G^{'}$ with the 3D gradient $\nabla G$, we
obtain Eq. \ref{eqn:r2}. The derivation of $b_v(k)$ requires the
two-point joint PDF and is more lengthy. We refer the readers
to DS10 for details. 

The velocity bias arises from correlation between ${\bf v}$ and
$\nabla \delta$.  Both the nonlinear evolution and smaller
smoothing scale $R_\Delta$ weaken such  correlation. We then expect weaker
deviation of velocity bias from unity, and therefore better agreement
with simulations.  Now we proceed to numerical evaluation using
Eq. \ref{eqn:r2} \& \ref{eqn:bv}.

\subsection{Numerical results under the  log-normal approximation}
The density field is known to be close to
log-normal \citep{1991MNRAS.248....1C,1994ApJ...420...44K,2000MNRAS.314...92T,2001ApJ...561...22K,2009ApJ...698L..90N}. Therefore to a good approximation, 
\ba
G(\delta)=\ln(1+\delta)-\langle \ln(1+\delta)\rangle\ .
\ea
Using the cumulant expansion theorem, we obtain
\ba
1+\delta&=&e^{G- \sigma_G^2/2},\ 
1+\sigma_\delta^2=\exp(\sigma_G^2)\ , \no \\
\sigma_{\nabla\delta}^2&=&\exp(\sigma_G^2)\sigma^2_{\nabla G}\ , \
\langle {\bf v}\cdot \nabla G\rangle=\langle {\bf v}\cdot \nabla \delta\rangle\ .
\ea
\bfi{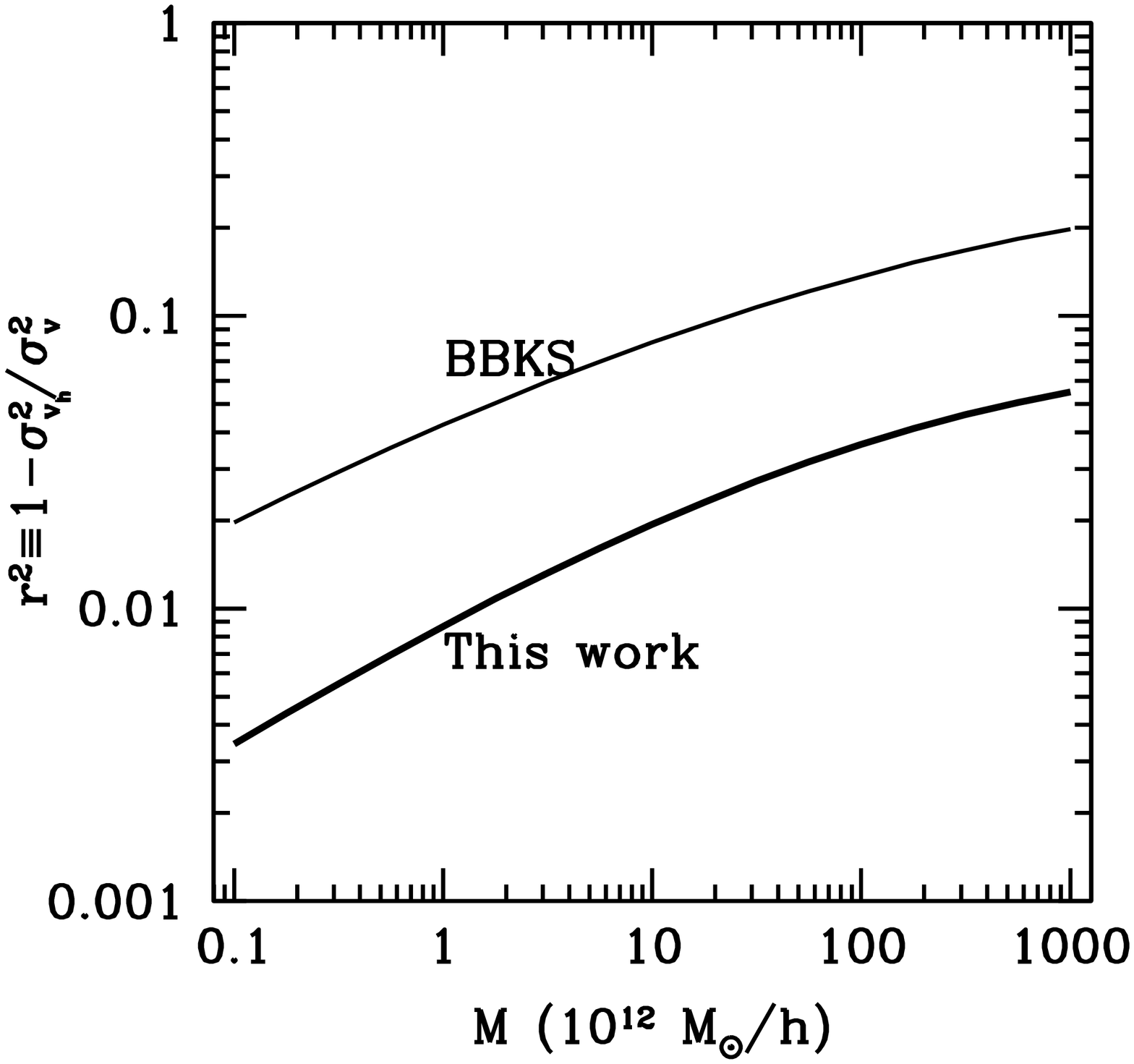}
\caption{The predicted difference between the $z=0$ halo velocity dispersion
  and matter velocity dispersion $1-\sigma^2_{v_h}/\sigma_v^2=r^2$, as
  a function of halo mass $M$. Our prediction is a factor of $\sim 4$
  smaller than the BBKS prediction. \label{fig:r2}}
\efi
Now $r$ and $R_v$ in Eq. \ref{eqn:r2} \& \ref{eqn:bv} can be expressed
with statistics of the density field, 
\ba
\label{eqn:r2Rv2}
r^2&=&\frac{\langle {\bf v}\cdot \nabla
  \delta\rangle^2}{\sigma_v^2\sigma^2_{\nabla\delta}}\left(1+\sigma_\delta^2\right)\
, \no\\ 
R_v^2&=&\frac{(1+\sigma_\delta^2)\ln(1+\sigma_\delta^2)}{\sigma^2_{\nabla
    \delta}}\ .
\ea
The corresponding properties above are determined by the nonlinear matter power
spectrum $P_\delta(k)$ through
\ba
\langle {\bf v}\cdot \nabla \delta\rangle&=&\int
\frac{k^3}{2\pi^2}P_{\theta\delta}(k)W^2_{\rm
  TH}(kR_{\Delta})\frac{dk}{k} \no \\
&=&\int
\frac{k^3}{2\pi^2}P_{\delta}(k) W^2_{\rm
  TH}(kR_{\Delta})\tilde{W}(k)\frac{dk}{k}\ , \no \\
\sigma_{\delta, \nabla \delta,v}^2&=&\int
\frac{k^3}{2\pi^2}P_{\delta}(k) W^2_{\rm
  TH}(kR_{\Delta}) k^{0,2,-2}\frac{dk}{k}\ .
\ea
$P_\delta$ is evaluated using the CAMB web
interface\footnote{https://lambda.gsfc.nasa.gov/toolbox/tb\_camb\_form.cfm},
which uses halofit \citep{Smith03} for the nonlinear power
spectrum. We adopt a flat $\Lambda$CDM cosmology with $\Omega_m=0.268$,
$\Omega_\Lambda=1-\Omega_m$, $\Omega_b=0.044$, $\sigma_8=0.83$,
$n_s=0.96$ and $h=0.71$.  $W_{\rm
  TH}(x)=3(\sin(x)-x\cos(x))/x^3$ is the top-hat window function.  The function $\tilde{W}(k)\leq 1$, introduced in \citet{Zhang13},
describes the impact of nonlinear evolution in the density-velocity
relation. We adopt the fitting formular in \citet{Zheng13} to
evaluate it. The nonlinear evolution weakens the density-velocity correlation, and
leads to weaker deviation of $b_v$ from unity. Including this effect
is also essential for correct prediction of halo 
velocity bias.

Numerical evaluations of Eq. \ref{eqn:r2} \& \ref{eqn:bv} at $z=0$ using
Eq. \ref{eqn:r2Rv2} are shown in Fig. \ref{fig:r2} \&
\ref{fig:bv}. The predicted $1-\sigma_{v_h}^2/\sigma_v^2=r^2$
increases with the halo mass. It is $0.3\%$, $0.9\%$,  $1.9\%$,
$3.6\%$ and $5.5\%$,  for halos of mass $10^{11,12,13,14,15}M_\odot/h$
respectively (Fig. \ref{fig:r2}). As a
comparison, the BBKS prediction is a factor of $4$ higher.  The
difference in $R_v^2$ is even larger (Fig. \ref{fig:bv}). Our model
predicts $R_v^2=0.098 ({\rm Mpc}/h)^2$ for $M=10^{12}M_\odot/h$. This
means that $1-b_v(k=0.1\hmpc)\simeq 0.1\%$. For $M=10^{13}M_\odot/h$
halos, $R_v^2=0.31 ({\rm Mpc}/h)^2$ and $1-b_v(k=0.1\hmpc)\simeq
0.3\%$. These predictions agree well with the finding of
$\mathcal{O}(0.1\%)$ deviation of $b_v$ from unity at $k\leq
0.1\hmpc$ \citep{Chen18}. For comparison, $R_v^2$ in DS10  is a factor of $10$-$20$
larger. 

\bfi{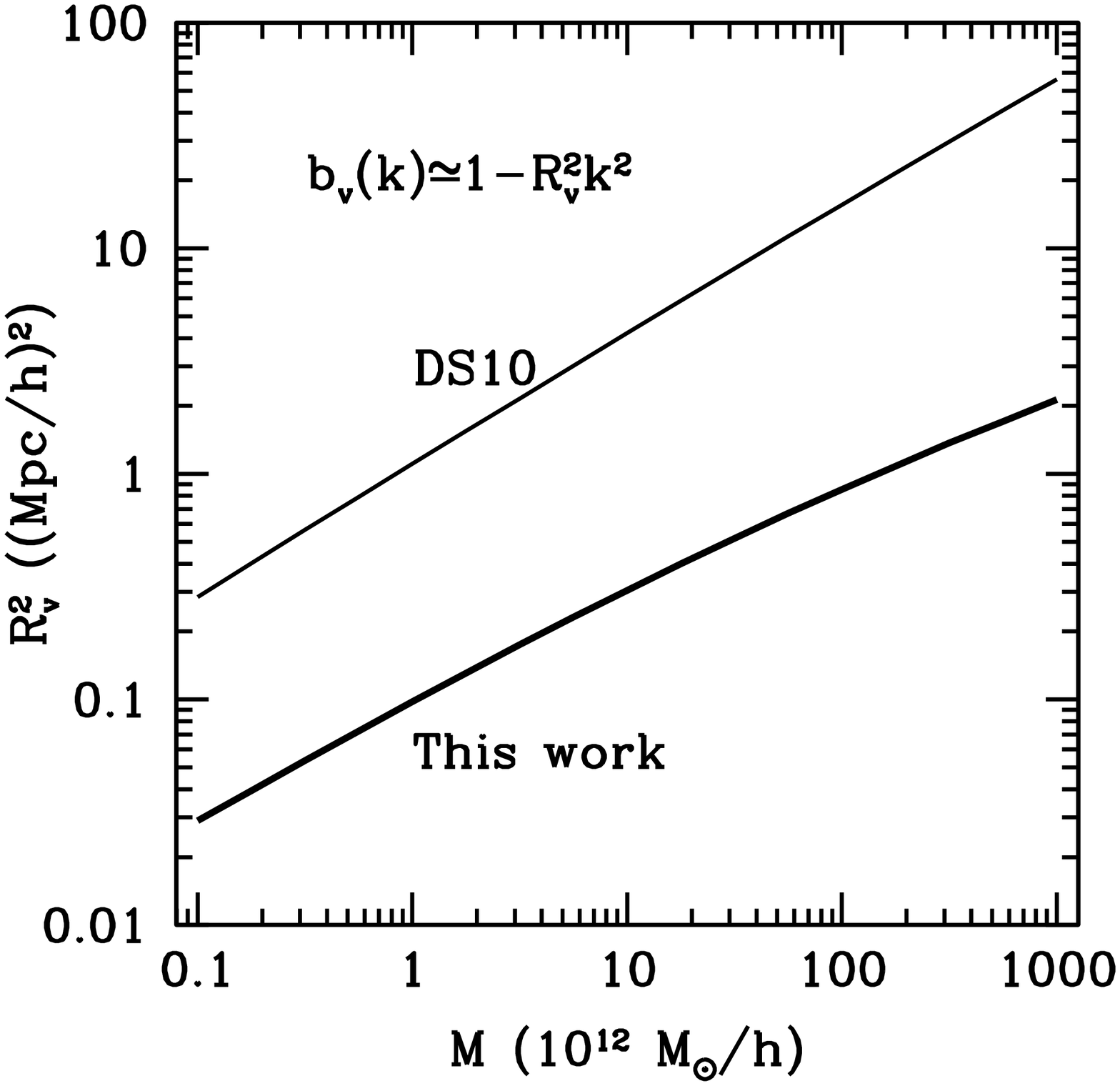}
\caption{The predicted $R_v^2$ ($b_v=1-R_v^2k^2$) at $z=0$, as a function of
  halo mass $M$. At $k=0.1\hmpc$, we predict $1-b_v\sim 0.1\%$,
  consistent with the recent simulation finding \citep{Chen18}. \label{fig:bv}}
\efi
\section{Discussions and conclusions}
Our model extends the peak model
methodology to nonlinear and non-Gaussian density field, and is
capable of  dealing with
halos instead of proto-halos.  The non-Gaussianity of the density
field, the smaller smoothing scale (halo virial radius $R_\Delta$
versus $R_S$)  and the weaker density-velocity correlation all have
impact on the velocity bias. The non-Gaussianity tends to amplify
  $r^2$ by a factor $1+\sigma_\delta^2$, and amplify $R_v^2$ by a
  factor $(1+\sigma_\delta^2)\ln(1+\sigma_\delta^2)/\sigma_\delta^2$
  (Eq. \ref{eqn:r2Rv2}), comparing to the 
  BBKS and DS10 expressions.  
  In contrast, the
  nonlinearity and  the smaller smoothing scale  suppress
  $r^2$ and $R_v^2$, by increasing $\sigma^2_{\nabla \delta}$ in the
  denominator (Eq. \ref{eqn:r2Rv2}). It further suppresses $r^2$
  through the $\tilde{W}<1$ factor in the numerator term $\langle {\bf
    v}\cdot \nabla\delta\rangle$.  Competitions of these opposite
  effect result in a weak deviation of $b_v$ from unity.\footnote{Interestingly, if we neglect the
  non-Gaussianity and non-linearity and simply replacing $R_S$ in the BBKS and
  DS10 expressions by $R_\Delta$, we also obtain weak deviation of
  $b_v$ from unity. But from the above discussions, this is a
  coincidence caused by the neglected non-Gaussianity and
  nonlinearity.}

Therefore we provide a feasible explanation of the vanishing volume
weighted halo velocity bias observed by \citet{Chen18}.  Nevertheless, our model may miss other necessary ingredients, since it does not explain all behaviors 
of velocity bias observed in simulations. First, it predicts incorrect
redshift dependence of velocity bias. \citet{Chen18} found that the
halo velocity bias monotonically increases with decreasing redshift,
regardless of halo mass. However, in our model the redshift dependence
of halo velocity bias is  not only weaker, but may also
be non-monotonic (for less massive halos). Second, it can not explain the observed
$b_v>1$ of $\la 10^{12}M_\odot$ halos at $z=0$. Our model, along with
BBKS and DS10, always predicts $b_v<1$. Both failures are likely
related to the imperfection of Gaussianization. Furthermore, approximating the
Gaussianization function with a lognormal transformation can result in
further error (Fig. 2, \citet{2009ApJ...698L..90N}).  Another possible
cause  is the subtlety in halo definitions.  The halo catalog \citep{Jing18} used by 
\citet{Chen18}  identifies halos with the Friends-of-Friends (FOF) algorithm of linking
length $b=0.2$. The corresponding $\Delta$ varies with the halo mass
(e.g. \citet{2011ApJS..195....4M}), while the virial overdensity
$\Delta$ adopted in our model is mass
independent. The two $\Delta$ also have different redshift
dependences. Furthermore, the \citet{Jing18} halo catalog excludes unbound particles in
halos after FOF. This further complicates the correspondence  between
halos in simulations and  in theory.  We are not able to address these
possibilities quantitatively and therefore leave them for future
investigation.

\section*{Acknowledgments}
This work was supported by the National Science Foundation of China
(11621303, 11433001, 11653003, 11320101002), and  National
Basic Research Program of China (2015CB85701).

\bibliographystyle{apj}
\bibliography{mybib}

\end{document}